# Missing Data and Prediction: The Pattern Mixture Kernel Submodel

**Sarah Fletcher Mercaldo\* and Jeffrey D. Blume, Ph.D.\*\***

Department of Biostatistics, Vanderbilt University, Nashville, TN

\**email:* sarah.fletcher@vanderbilt.edu

\*\**email:* j.blume@vanderbilt.edu

SUMMARY:   Missing data are a common problem for both the construction and implementation of a prediction algorithm. Pattern mixture kernel submodels (PMKS) - a series of submodels for every missing data pattern that are fit using only data from that pattern - are a computationally efficient remedy for both stages. Here we show that PMKS yield the most predictive algorithm among all standard missing data strategies. Specifically, we show that the expected loss of a forecasting algorithm is minimized when each pattern-specific loss is minimized. Simulations and a re-analysis of the SUPPORT study confirms that PMKS generally outperforms zero-imputation, mean-imputation, complete-case analysis, complete-case submodels, and even multiple imputation (MI). The degree of improvement is highly dependent on the missingness mechanism and the effect size of missing predictors. When the data are Missing at Random (MAR) MI can yield comparable forecasting performance but generally requires a larger computational cost. We see that predictions from the PMKS are equivalent to the limiting predictions for a MI procedure that uses a mean model dependent on missingness indicators (the MIMI model). Consequently, the MIMI model can be used to assess the MAR assumption in practice. The focus of this paper is on out-of-sample prediction behavior; implications for model inference are only briefly explored.

## 1. Introduction

### 1.1 *The Problem*

While missing data are problematic for both estimation and prediction, the statistical literature has been largely focused on addressing the impact of missing data on estimation procedures and parameter inference. Missing data present a two-fold problem for forecasting: first, in building a model, and second, in using the model to make out-of-sample predictions for individuals with missing predictors. Here we focus on the second problem, specifically evaluating what Wood et al. (2015) defines as Pragmatic Model Performance, which refers to the model's performance in a future clinical setting where some individuals may have partly missing predictors.

It is often assumed that imputation methods, because they improve parameter estimation procedures, also improve out-of-sample prediction performance. However, this is just speculation, and our investigation indicates that this is the exception rather than the rule. Ideally, it would be possible to find a prediction rule that uses simple imputation or did not require imputation, and thus was much less computationally burdensome and more readily applied in practice. The impact of missing data for out-of-sample prediction is uniformly underestimated. A poor imputation algorithm at the prediction stage can drastically reduce a model's overall prediction performance, and data frequently go missing at this stage in real life.

### 1.2 *Current Approaches to Imputation*

Typical strategies for dealing with missing predictors are driven by practical constraints. Common strategies include zero imputation and mean imputation, which are trivial to implement, but often lead to poor predictions. Conditional mean imputation and multiple imputation can be implemented with accessible software when fitting a model, but they are rarely used in the clinic when predictors are missing for out-of-sample predictions (Janssen et al., 2009). The advantages and drawbacks of imputation methods for out-of-sample imputation procedures are listed in Table 1. The obvious issue, not well addressed in the literature, is the extent to which these approaches degrade prediction performance, as later shown.

[Table 1 about here.]

Multiple Imputation (MI) draws multiple placeholder values from conditional distributions derived from the observed data (van Buuren, 2012; Janssen et al., 2010; Harrell, 2013), uses the placeholder values to fit the model, and then combines the models using Rubin's rules (Rubin, 2009). When the data are missing at random (MAR), MI can substantially increase inferential efficiency by leveraging information in incomplete data records. The 'best' predictions from a multiply imputed prediction model are the model's predictions averaged over all imputation sets (Vergouwe et al., 2010; Wood et al., 2015). Recently, the popularity of MI as the primary 'principled' strategy for constructing and applying a prediction model in the presence of missing data has grown (Harrell, 2013; Janssen et al., 2009).

However, applying a multiply imputed prediction model to an out-of-sample individual who is missing predictor information is not straightforward. This is because, technically speaking, the predictions need to be re-estimated with the imputation procedure based on the original data and the new out-of-sample record if you want to apply Predictive Mean Matching or K-Nearest Neighbor imputation techniques (as we did in our simulatons and examples). Of course, this requires the original data, the imputation datasets, and substantial on-demand computing power, which is often impractical in real world settings. Moreover, this approach often has a heavy computational burden and is not easily programmed in web applications (because the imputation algorithm must be repeated for every out-of-sample prediction). One could ignore this step and use the multiply-imputed model along with some one-step imputation procedure using the saved chain equations or fitted conditional distributions, but this process is not likely to be congenial with the original fitting approach.

### 1.3 *Proposed Solution*

Here we propose using an approach that we call the Pattern Mixture Kernel Submodels (PMKS) procedure. PMKS postulates a unique prediction model for each missing data pattern and estimates that model using only the subjects in that pattern. As a result, PMKS requires no imputation algorithm. Details are provided in section 3.1. As a forecasting algorithm, PMKS benefits greatly from the reduction in prediction bias that comes with the pattern-specific approach. The loss in efficiency that can results when the data are MAR, or when the patterns have common parameters, is often small in comparison. Moreover, we will see the PMKS is optimal in the sense of minimizing the expected prediction loss. Because of these advantages, we anticipate that PMKS will have broad impact in the arena of big data where prediction is paramount and MI is often computationally unfeasible e.g., when using an entire system of electronic medical records.

### 1.4 *Organization*

Section 2 defines our notation and provides a brief background to key missing data concepts. Section 3 describes our proposed methods, provides a simple example, and draws connections between PMKS and MI models in some generality. Section 4 describes extensive simulations of PMKS in order to establish that wide range of setting under which PMKS excels. Section 5 describes the performance of PMKS compared to other imputation strategies applied to the SUPPORT Study, a multi-center two phase study of 9105 patients, from which a day 3 Physiology Score (SPS) was predicted (Phillips et al., 1995). Section 6 provides some brief concluding remarks.

## 2. Notation and Background

### 2.1 *Notation*

Let $\boldsymbol{Y} = (Y_1, ..., Y_n)$ be the vector of $n$-length observed responses. With our focus on prediction, we assume all responses are observed. This assumption can be relaxed, but it is not necessary to our discussion here. Predictors (covariates) are denoted by a $(n \times p)$ matrix $\boldsymbol{X} = (\boldsymbol{X_1}, ..., \boldsymbol{X_p})$ where $\boldsymbol{X_j} = (X_{1j}, ..., X_{nj})^T$ for $j = 1, ..., p$ predictor vectors of length $n$. Let $\boldsymbol{M} = \{M_{ij}\}$ be the $(n \times p)$ matrix of missing data indicators where indicator $M_{ij} = 1$ if $X_{ij}$ is missing and $M_{ij} = 0$ if $X_{ij}$ is observed for $i = 1, ..., n$ individuals and $j = 1, ..., p$ parameters.

To differentiate between models, we will use different greek symbols for their parameters. For example, parameters in the pattern mixture kernel submodels (PMKS) will be donated with a $\boldsymbol{\gamma}$. Parameters in a traditional regression mean model $E[Y|\boldsymbol{X}] = \boldsymbol{X}\boldsymbol{\beta}$ - those typically of interest in an estimation setting - will be donated by $\boldsymbol{\beta}$. Notice this is a strong assumption with regards to the missing data, forcing the same mean-response function for every missing data pattern, and implying a MAR mechanism within a single marginal model. Parameters representing the effects of the missingness indicators, $\boldsymbol{M}$, will be denoted by $\boldsymbol{\delta}$; these parameters will distinguish our MIMI model (defined in section 3.5) from a traditional MI model.

### 2.2 *Pattern Mixture and Selection Models*

Our approach has roots in the established literature on pattern mixture models (Little (1993)). The traditional pattern mixture model factorization is:

$P(\boldsymbol{Y}, \boldsymbol{M}|\boldsymbol{X}, \boldsymbol{\gamma}, \boldsymbol{\pi}) = P(\boldsymbol{Y}|\boldsymbol{X}, \boldsymbol{M}, \boldsymbol{\gamma})P(\boldsymbol{M}|\boldsymbol{X}, \boldsymbol{\pi})$, where $\boldsymbol{\pi}$ is a parameter vector for the missingness mechanism (Little and Rubin (2014)). The pattern-mixture approach allows for a different response model in each missing data pattern. An alternative formulation is the selection model: $P(\boldsymbol{Y}, \boldsymbol{M}|\boldsymbol{X}, \boldsymbol{\theta}, \boldsymbol{\omega}) = P(\boldsymbol{Y}|\boldsymbol{X}, \boldsymbol{\theta}, \boldsymbol{\omega})P(\boldsymbol{M}|\boldsymbol{Y}, \boldsymbol{X}, \boldsymbol{\omega})$ where $\boldsymbol{\theta}$ and $\boldsymbol{\omega}$ are

parameter vectors (Little and Rubin (2014) Little and Wang (1996)). This factorization describes the (single) marginal response model. In this paper, we will not explicitly consider selection models except to use them to generate data from certain missing data mechanisms that cannot be simulated otherwise. While the selection model allows for $\boldsymbol{Y}$ and $\boldsymbol{M}$ to be either independent or dependent, the pattern-mixture model is used when the response model changes by missing data pattern. This flexibility, it turns out, lends an advantage to our proposed PMKS prediction algorithm.

### 2.3 *Missingness Mechanisms*

To describe missing data, we use the following missingness mechanisms: Missing Completely at Random (MCAR), Missing at Random (MAR), Missing Not at Random (MNAR), Missing at Random where the missingness depends on $Y$ (MARY), and Missing Not at Random where the missingness depends on $Y$ (MNARY) (Little and Rubin (2014)). The latter two mechanisms can only be simulated in the selection model formulation. A more detailed description of these missingness mechanisms is given later in section 3.

If the missingness mechanism is MCAR, then pattern mixture and selection models are equivalent (Little, 1993). When the data are not MCAR, the parameters of the kernel functions associated with the selection and pattern mixture models have different interpretations and care must be taken when estimating and interpreting them. The selection model describes the marginal relationship of $\boldsymbol{Y}$ on $\boldsymbol{X}$ while the pattern mixture model describes the relationship of $\boldsymbol{Y}$ on $\boldsymbol{X}$ conditional on $\boldsymbol{M}$. Marginal effects from the selection model are generally not identifiable in the context of a pattern mixture model, although some parameterizations can be identified though complete case restrictions that essentially force equality restraints on certain parameters (Little, 1993). Identifiability is obviously a problem when the goal is estimation and data are MNAR. However, here our goal is prediction and complex re-parameterizations of marginal effects are not a major impediment even if the mapping is not easily reversed. If one marginal model is truly of interest, it is always possible to marginalize over the pattern-specific model. Of course, how that model should be interpreted when the data are not MAR is not immediately clear.

## 2.4 *Complete Case Models and Submodels*

Two alternatives to multiple imputation are complete case models and complete case submodels. A complete case analysis simply ignores the records with missing data and estimates a single model. Note that complete-case models have routinely poor prediction performance when the data are not Missing Completely at Random (MCAR) (Knol et al., 2010; Janssen et al., 2009). Complete Case Submodels (CCS) use all available data to fit a submodel for each missing data pattern. That means, for CCS, a data records often contribute to multiple patterns/submodels. This approach is vastly different from a complete-case analysis where just a single model is fit using only subjects with complete data.

Note that the complete case model does not solve the problem with missing out-of-sample predictors; some type of imputation is still required. In contrast, CCS do not have this problem. Their predictions come from the submodel that matches the out-of-sample missing data pattern and so no imputation is needed. For PMKS, each data record contributes only to a single pattern/submodel, which turns out to be a key difference compared to CCS.

## 3. Methods

### 3.1 *Pattern Mixture Kernel Submodels*

PMKS has roots in pattern mixture model, using the kernel of the pattern mixture model as a prediction machine. These pattern-specific models are the PMKS. To make predictions, we fit the set of models $\{\hat{f}_1, ..., \hat{f}_k\}$ where $\hat{f}_m = \hat{f}_m(\boldsymbol{X}, \boldsymbol{M})$ is the pattern mixture model in pattern $m = 1, ..., k$ where $k \leqslant 2^p$ different patterns. Here we consider straightforward prediction algorithms such as $\hat{f}_m = E(\boldsymbol{Y}|\boldsymbol{X}, \boldsymbol{M}; \hat{\boldsymbol{\gamma}}_m)$ where $\hat{\boldsymbol{\gamma}}_m$ is the vector of estimated

pattern specific parameters. However, our findings apply to any generalized linear model or machine learning prediction technique. Although up to $k = 2^p$ different models might have to be fit, in practice only a small fraction of those patterns are observed.

Each PMKS is fit using only subjects from that pattern. This is in contrast to Complete Case Submodels (CCS) where the submodels are fit using all available data. This subtle difference turns out to be critically important; as it is only appropriate under certain missing data mechanisms. For comparison, we denote the set of CCS models as $\{\hat{g}_1, ..., \hat{g}_k\}$ where $\hat{g}_m = \hat{g}_m(\boldsymbol{X}) = E(\boldsymbol{Y}|\boldsymbol{X}; \hat{\boldsymbol{\beta}}^*_m)$ for patterns $m = 1, ..., k$. Here we use $\hat{\boldsymbol{\beta}}^*_m$ as the estimated pattern specific parameter vector. The asterisk here is used to distinguish the CCS parameters from those obtained in a complete case analysis ($\hat{\beta}$). The difference between $\hat{f}_m$ and $\hat{g}_m$ is illustrated below in the context of a simple example. Later in section 3.4, we discuss how to fit PMKS and CCS when a pattern is not observed or the data are too sparse. CCS have the advantage of being fittable in many of these situations with the obvious drawback of its strong dependence on the MAR assumption.

### 3.2 *A Simple Example*

A simple illustration helps to fix ideas. Consider a linear model for continuous outcome $Y$ with two covariates $X_1$, $X_2$. There are only four missing data patterns: (1) $X_1$, $X_2$ both observed; (2) $X_1$ missing, $X_2$ observed; (3) $X_1$ observed, $X_2$ missing; (4) $X_1$, $X_2$ both missing. The pattern mixture kernel submodels are the set of corresponding response models in each missing data pattern. These are given in table 2.

[Table 2 about here.]

Using the above notation, the estimated PMKS response function for $E[Y|X_1, M_1 = 1, M_2 = 0]$ is $\hat{f}_3$ while the CCS analogue for $E[Y|X_1]$ is $\hat{g}_3$. Note that $\gamma_{p,m}$ does not necessarily equal $\beta_p^*$ for any $m = 1, 2, 3, 4$. Also, none these submodels corresponds to the typical marginal model expressed as $E[Y|X_1, X_2] = \beta_0 + \beta_1 X_1 + \beta_2 X_2$ where the parameters $\boldsymbol{\beta}$ represent traditional direct effects. It is tempting to assume that the model based on pattern 1 would yield proper estimates of this marginal model, but this only happens when the data are MAR on the covariates (and not on the response). This is because that is the only case where the marginal model corresponds exactly with the data generating mechanism in each pattern. That is, it is the only case where the marginal model is true mean response model for every pattern (White and Carlin (2010); Bartlett et al. (2014)).

### 3.3 *Prediction Performance of PMKS*

PMKS is computationally efficient for missing data because it avoids the issue; it fits a series of models in which none have any missing data. Fitting each submodel is now straightforward because the missing data problem has been avoided. The key realization is that minimizing the expected loss in each pattern amounts to minimizing the expected loss marginally. Thus, we need only use standard techniques to fit and cross-validate the pattern specific models.

### 3.3.1 *Minimizing the Expected Prediction Error.*

Minimizing the expected prediction error in each pattern will, in turn, minimize the overall expected prediction error. To see this, note that:

$$E_{Y|X}[L(Y, \hat{f}(\boldsymbol{X}))] = E_M \left[ E_{Y|\boldsymbol{X},\boldsymbol{M}} \left[ L\left(Y, \hat{f}_m\right) \right] \right]$$
$$= \sum_M P(M) E_{Y|\boldsymbol{X},\boldsymbol{M}} \left[ L\left(Y, \hat{f}(\boldsymbol{X}, \boldsymbol{M})\right) \right]$$

where $\hat{f}_m = \hat{f}_m(\boldsymbol{X}, \boldsymbol{M})$. Hence, selecting $\hat{f}_m$ to minimize the pattern specific expected loss, $E_{Y|X,M} \left[ L\left(Y, \hat{f}_m(\boldsymbol{X}, \boldsymbol{M})\right) \right]$, will in turn minimize the overall loss $E_{Y|X}[L(Y, \hat{f}(\boldsymbol{X}))]$.

Here $L\left(Y, \hat{f}_m(\boldsymbol{X}, \boldsymbol{M})\right)$ must be a properly defined pattern-specific loss function. The loss function is flexible; it could be squared error or 0/1 loss (Hastie et al., 2009). But the pattern-specific restriction is important; the result might not hold for certain metrics where predictions in one pattern are compared with predictions in another (for example, the area under the ROC curve). In that case, the overall AUC is not equal to the average of pattern specific AUCs, which is the property we need in order for to take advantage of this approach. Fortunately, most common loss functions of interest in prediction problems have this property.

Note that constructing a prediction model within each missing data pattern effectively resolves the missing data dilemma because the missing predictors are missing for everyone in that pattern and only marginal effects can be estimated. The result implies that, in practice, prediction models should be constructed and cross validated within each pattern in order to maximize predictive ability. The only reason to do this marginally is if the MAR assumption is known to hold. Then direct estimation of $E_{Y|\boldsymbol{X}}\left[Y, \hat{f}(\boldsymbol{X})\right]$ is complex, in part, because only a single model $\hat{f}$ is used for all predictions and fitting that model with missing data requires a complex algorithm such as multiple imputation or the EM algorithm to handle the missing data. Since each PMKS is directly estimable with routine tools, the right side of the equation can be fit, minimized, and cross validated rather easily. This simple argument impacts practice because $M$ is often ignored in the modeling stage due to historical concerns about the missing indicator method.

3.3.2 *PMKS loss is a weighted average of a full and reduced model.* For a linear model the squared prediction error is a common and relevant loss function. To examine the bias-variance tradeoff in PMKS, it is helpful to revisit a simple example given by Shmueli (2010) in which the Expected Prediction Error (EPE) is evaluated for a "fully specified model" (large) versus an "underspecified model" (small). Suppose data come from the model $f(x) = \beta_0 + \beta_1 x_1 + \beta_2 x_2 + \epsilon$ with $\epsilon \sim N(0, 1)$. When no predictors are missing we estimate the full model as $\hat{f}(x) = \hat{\beta}_0 + \hat{\beta}_1 x_1 + \hat{\beta}_2 x_2$. Here the expected prediction error (EPE) is the sum of the bias, variance, and irreducible error of the predictions or fitted values (Hastie et al. (2009)):

$$\text{EPE}_L = E\left[\left(Y - \hat{f}(x_1, x_2)\right)^2\right] = \sigma^2\left(1 + [1 \quad x_1 \quad x_2](X_L'X_L)[1 \quad x_1 \quad x_2]'\right)$$

where $\text{EPE}_L$ denotes the EPE of the full model. In contrast the EPE of the underspecified or submodel is given by:

$$\text{EPE}_S = E\left[\left(Y - \hat{f}^*(x_1)\right)^2\right] = ((\gamma_0 + x_1\gamma_1) - (\beta_0 + \beta_1 x_1 + \beta_2 x_2))^2 + \sigma^2[1 \quad x_1](X_S'X_S)[1 \quad x_1]'$$

where $\hat{f}^*(x) = \hat{\beta^*}_{0,1} + \hat{\beta^*}_{1,1}x_1$. Note that in this case $\hat{f}^*(x) = \hat{g}_2$ . The EPE of the PMKS model is just a weighted average of the large and small prediction models.

$$\text{EPE}_{PMKS} = \sum_m P(M = m)\text{EPE}_m = \text{EPE}_L(1 - P(M)) + \text{EPE}_S P(M)$$

To illustrate the bias variance tradeoff, we simulated the EPE in figure 1. The simulation fixes $X_1 = x_1$, and draws from the conditional distribution $X_2|X_1 = x_1 \sim N(\mu_2 + \frac{\sigma_2}{\sigma_1}\rho_{1,2}(x_1 = \mu_1), (1 - \rho_{1,2}^2)\sigma_2^2)$. The EPE for the correctly specified full model is just the irreducible error, whereas the EPE for the underspecified model increases as the out-of-sample predictor moves away from its population mean.

[Figure 1 about here.]

The out-of-sample prediction error from the large model is given by the green line in figure 1, and is equal to the model variance. If the data were generated from the large model and predictions were given from the small model that includes only $X_2$, then the expected prediction error is approximated by the purple points in figure 1.

The yellow points in figure 1 denote the prediction error that arised from the PMKS in this setting; $\hat{f}_1$ makes predictions when all data are available and $\hat{f}_2$ makes predictions when only $X_2$ is available. Clearly, PMKS has smaller EPE for every out-of-sample $X_1$. In this case the probability of missingness was 50%, $P(M_1 = 1) = 0.5$.

### 3.4  *Practical Considerations*

As discussed earlier, multiple imputation has a substantial computational burden for out-of-sample predictions with missing data because the imputation algorithm must be repeated adding the new person in the data run. PMKS, on the other hand, do not need to be re-computed for every new prediction. The upfront computational effort is large for $2^p$ patterns, but it is often minor compared to the MI machinery required in the same setting and in practice only a fraction of available patterns are observed. When data are sparse within a pattern, it may not be possible to fit the PMKS. In such cases it is necessary to make assumptions and the CCS approach is a reasonable option. This hybrid approach has worked well for us in practice, in large part because the contribution to the EPE for patterns that are too sparse to fit with PMKS is often negligible.

If $p$ is very large, and storing $2^p$ prediction models seems unreasonable, there are several options. First, fit models only for observable patterns, ignoring patterns not observed. Second, only fit models for patterns in which the missing variable, or combinations of missing variables, are 'important' to the predictions. Third, if the data are available in real time, it may be possible to fit PMKS on demand, since imputation is not necessary, and this reduces the need to store all $2^p$ models simultaneously. Lastly, shrinkage methods to the MIMI model, discussed in section 3.5, can indicate how best to borrow strength over the patterns.

It is important to distinguish between the computation cost between the in-sample model construction phase and out-of-sample prediction phase. Both PMKS and MI could have high in-sample computation cost depending on the number of predictors and the data size. But as described in table 1, the out-of-sample computational costs for PMKS are negligible, whereas for MI they can remain intensive, even for a single individual. Importantly, PMKS does not require missing data mechanisms to be consistent in the data used to construct the model and the target population. This is because, conditional on the missingness pattern, the data are effectively MAR. The MAR assumption implies $(M \perp X_j)|\boldsymbol{X_{-j}}, \forall j = 1..p$. PMKS reformulates this assumption as $(M \perp X_j)|(\boldsymbol{X_{-j}M_{-j}})$, allowing the MAR assumption to hold within each pattern (Little, 1993).

### 3.5 *PMKS as the limit of MIMI model*

There are some interesting connections between PMKS and MI. PMKS is the limit of a congenial MI procedure when the mean model depends on the missing data indicators. This new MI procedure - which we call this the Multiple Imputation with Missingness Indicators (MIMI) model - can be used to assess the MAR assumption in practice. The MIMI model also makes the context clear about what elements of the model can be assumed identical across the patterns for inference purposes. The utility of missingness indicators can only be realized through an imputation procedure, an often overlooked and important point. The implications for estimation will be discussed elsewhere, as the focus of this paper is on prediction, but the connection is an important one.

The MIMI model is a multiple imputation model that is dependent on the indicators $M_i$ from $i = 1, .., p$. Typically, the mean model would depend on the missingness indications. For example, in the past example with $p = 2$ covariates, $X_1$ and $X_2$, we might consider the following mean model:

$$\begin{aligned}
E[Y|X_1, X_2, M_1, M_2] = \beta_0 + \beta_1 X_1 + \beta_2 X_2 + \delta_1 M_1 + \delta_2 M_2 \\
+ \delta_3 X_1 M_1 + \delta_4 X_2 M_2 + \delta_5 X_1 M_2 + \delta_6 X_2 M_1
\end{aligned} \tag{1}$$

where the $\boldsymbol{\beta}$ parameters represent the traditional direct effects of interest and the $\boldsymbol{\delta}$ parameters, which we will call auxiliary parameters, explain how the traditional effects change by missing data pattern. If the data are MCAR, then $\delta_i = 0 \ \forall \ i$. Otherwise, the traditional effects might not exist as the dependencies in the model parameters can be complex. An informal test of the $\boldsymbol{\delta}$ parameters may provide some insight into the observed missing data mechanism. If there is no evidence to suggest that the indicators contribute to the model predictions, this may help to make the case for a MAR mechanism. Shrinkage methods may also be applied to the delta $\boldsymbol{\delta}$ parameters to help assess which covariates have non-ignorable missing data mechanisms, and will be addressed in future work.

Note, Molenberghs et al., 2008 describes a longitudinal setting where every MNAR model can be decomposed into a set of MAR models. Molenberghs et al., 2008 rightly asserts that this duality is problematic for inference about a parameter. However, as noted in the paper, these representations will yield different predictions and so the implications are different in out context where prediction is the objective. The results from Molenberghs et al., 2008 are important in that they aligns with our results in the non-longitudinal setting, were predictions must match those from the natural pattern mixture model (that is MAR by definition conditional on the pattern) in order to retain its predictive optimality.

Unfortunately, model (1) cannot be properly fit unless the missing predictors are imputed. When the missing data are imputed with a proper MI algorithm, the coefficients of (1) are essentially identified by that algorithms imputation scheme and the auxiliary parameters become estimable under those assumptions. Model (1) is an example of the Multiple Imputation with Missingness Indicators (MIMI) model. Note that MIMI makes certain assumptions about the missingness mechanism, and these assumptions will lead to different fits of the auxiliary parameters. This flexibility is both good and bad; good because we could use the auxiliary parameters to check assumptions, and bad because the auxiliary parameters are inherently unidentifiable.

Adding missingness indicators to a model (when the goal was parameter estimation) has received criticism historically. The simply plug-in varieties of the missing-indicator method yields biased parameter estimates even in simple cases where data are missing completely at random (MCAR) (Allison, 2001; Groenwold et al., 2012). The classical missing-indicator method fills in a constant (often zero or the overall mean) for the missing values and augments the data design matrix with a binary indicator for each covariate with missing values. However, when the missing-indicator method is combined with proper imputation methods the model produces unbiased parameter estimates in the same cases in which complete case estimation is unbiased (Jones, 1996; Dardanoni et al., 2011, 2015). That is, when $M_i$ and $Y$ are conditionally independent given $X_i$, then for any choice of imputation matrix, the OLS estimate of (1) coincides with the OLS estimate of $\boldsymbol{\beta}$ in the complete case model (Bartlett et al., 2014; Jones, 1996; Dardanoni et al., 2011; White and Carlin, 2010). Thus, the MIMI model is essentially an extension of the ideas in (Jones, 1996) to a more flexible multiple imputation setting.

Although the missing-indicator method has been heavily investigated in the context of inference, this method has not been explored for prediction where a bias-variance tradeoff may be more desirable. The MIMI model leverages multiple imputation to create placeholders for the missing data that do not negatively impact the models predictive ability. But of course, the value of these imputations does impact the estimation of direct effects and the properties of those estimators.

The connection between the MIMI model and PMKS can be seen through a simple rearrangement of the mean model (1). There are differences; the MIMI model forces constant variance across all missing data patterns, whereas PMKS allows the variance to change by pattern. PMKS are most easily understood as projections of the true pattern-specific model into the space of observed covariates. As such, slope parameters for observed covariates may be distorted if missing covariates are correlated with observed covariates. In those cases, the model is really estimating the total effect when the direct effect is the quantity of interest. PMKS is a series of models based on the total effects that can be estimated from the data at hand, while MIMI tries to reparametrize each patter-specific mean model and average the direct effects of interests.

Applying the plug-in principle, the MIMI mean model reduces to the PMKS when conditional mean imputation is used to impute missing covariates. Denote the imputed covariates as $X_i^* = E[X_1|X_2] = \alpha_0 + \alpha_2 X_2$ if $X_{i1}$ is missing and $X_{i1}$ otherwise.

Rearranging the MIMI model we have:

$$
\begin{aligned}
E[Y|X_1, X_2, M_1, M_2] = &(\beta_0 + \delta_1 M_1 + \delta_2 M_2) \\
&(\beta_1 + \delta_3 M_1 + \delta_5 M_2)X_1 \\
&(\beta_2 + \delta_4 M_2 + \delta_6 M_1)X_2
\end{aligned}
$$

Which just reduces to PMKS. To illustrate, for the 4 patterns in our running example:

$$
\begin{aligned}
E[Y|X_1, X_2, M_1 = 0, M_2 = 0] &= \beta_0 + \beta_1 X_1 + \beta_2 X_2 \\
E[Y|X_1, X_2, M_1 = 1, M_2 = 0] &= (\beta_0 + \delta_1) + (\beta_1 + \delta_3)E[X_1|X_2] + (\beta_2 + \delta_6)X_2 \\
&= (\beta_0 + \delta_1) + (\beta_1 + \delta_3)(\alpha_0 + \alpha_2 X_2) + (\beta_2 + \delta_6)X_2 \\
&= (\beta_0 + \delta_1 + \beta_1\alpha_0 + \delta_3\alpha_0) + (\beta_2 + \delta_6 + \beta_1\alpha_2 + \delta_3\alpha_2)X_2 \\
&= \gamma_0 + \gamma_2 X_2
\end{aligned}
$$

Note that model $E[Y|X_1, X_2, M_1 = 1, M_2 = 0] = \gamma_0 + \gamma_2 X_2$ is the submodel including only the covariate $X_2$ fit within the group of individuals who are missing the covariate $X_1$ (this is why the conditioning on $M$ is important) and is equivalent to $\hat{f}_2$ in section 3.1. Hence, PMKS and MIMI are two parameterizations of the 'same' model. To examine this, we next present simulations in the linear model case under a wide variety of missing data mechanisms. This principle extends to the GLM setting, on the linear predictor scale, and is currently under investigation by the authors. More complex prediction machines that are highly non-linear should exhibit the same general behavior.

## 4. Simulations

We generated $n$ multivariate normal predictor vectors according to $\left( \begin{smallmatrix} x_1 \\ x_2 \end{smallmatrix} \right) \sim N(\boldsymbol{\mu}, \boldsymbol{\Sigma})$, where $\boldsymbol{\mu} = (3,3)$ and $\boldsymbol{\Sigma} = \left( \begin{smallmatrix} 1 & 0.5 \\ 0.5 & 1 \end{smallmatrix} \right)$, for example, are set to provide certain predictor profiles in terms of their correlation. Simulated outcomes $Y$ are generated from various combinations of $x_1$ and $x_2$. The pattern mixture model formulation uses $\boldsymbol{X}$ to induce one of three missing data mechanisms, MCAR, MAR, or MNAR. The outcome $Y$ is then generated from the MIMI mean model using the true $\boldsymbol{X}$ values and the simulated missing data indicators. Here the missingness may only depend on the predictors vector $X$. In contrast, the selection model formulation simulates $Y$ from the marginal model $Y = \beta_0 + \beta_1 X_1 + \beta_2 X_2 + \epsilon$, where $\epsilon \sim N(0,1)$. Missing data indictors are then induced according to the desired mechanism. Note that here the missingness may depend on the outcome $Y$. A more complex model can always be reduced to a linear combination of non-missing variables, and missing variables, and so this simple example is representative of more complicated situations.

We simulated the following five missing data mechanisms as defined in table 3 for this situation: MCAR, MAR, MNAR, MARY, and MNARY. The latter two mechanisms could only be simulated in the selection model formulation. We forced the missingness data mechanism to be consistent between the in-sample and out-of-sample populations, and $\nu_0$ is empirically calculated to maintain the desired probability of missingness.

[Table 3 about here.]

### 4.1 *Parameters*

Parameters profiles explored were $\beta_1 = 1, 3, 5$, $\rho = 0, 0.5, 0.75$, $P(M_1 = 1) = 0.20, 0.50, 0.75$, and $n = 50, 200, 500, 1000$. We present here only one case that was largely representative of our findings: $\beta_1 = 3$, $\rho = 0.5$, $P(M_1 = 1) = 0.50$, and $n = 1000$. For the out of sample population we assumed one-by-one enrollment. Missing data was impute by zero imputation, unconditional mean imputation, single conditional mean imputation using a Bayesian conditional mean model, single conditional mean imputation using a frequentist conditional mean model, or multiple imputation (predictive mean matching, 10 imputations). We fixed the imputation engine based on the in-sample population to closely mimic real world application of these methods.

### 4.2 *Simulation procedure*

We compared the performance of PMKS, complete case model predictions, complete case submodel (CCS) predictions, traditional MI and the MIMI imputation model. The full simulation procedure was as follows: (1) data are generated and missing data indicators are generated according to the missing data mechanism in table 3, as described in section 5.1; (2) missing data are imputed; (3) the MI model, MIMI model, CCS, and PMKS models are fit; (4) step 1 is repeated to obtain a new out-of-sample population; (5) individuals are imputed one by one, using the above imputation procedures, assuming a fixed imputation engine from the in-sample population; (6) individual predictions and performance measures are computed; (7) steps 1 through 6 are repeated 10,000 times.

A squared error loss function was used to compare performance of the approaches. For example the squared error loss across all missing data patterns in the PMKS is $\frac{1}{n} \sum_i \sum_j P(M_i = 1)(Y_{ij} - \hat{Y}_{ij})^2$ where $j = 1, , n$ subjects and $i = 1, , m$ patterns. This loss is the averaged over the 10,000 simulations to approximate the expected loss. Table 4 shows the average squared imputation error for predictor $x_1$ as a function of imputation strategy and missingness mechanism.

### 4.3 *Simulation Results*

Results are presented for the following set of parameters: $\beta_0 = 1, \beta_1 = 3, \beta_2 = 1, \delta_1 = 1, \delta_3 = 1, P(M_1 = 1) = 0.5, \nu_1 = 1, \nu_2 = 1, \nu_{1,Y} = 1, \nu_{2,Y} = 1$. There were negligible differences in pattern specific and total squared error loss for the MCAR missing data mechanism. For all missing data scenarios, MI and conditional mean imputations resulted in a biased parameter estimation. This bias appears most clearly in predictions for observations without missing data (blue dots). When $Y$ is added to the MI model, the model parameters had negligible bias. However, since the out-of-sample $Y$ is missing, the out-of-sample imputations of $x_1$ have greater bias than the imputation model in which $Y$ is not included resulting in a higher total prediction error (e.g., see table 4).

When $Y$ is generated from a selection model formulation, all methods perform similarly (apart from MI as described above) under the MAR missing data mechanism. When data are MNAR, PMKS and the MIMI models have slightly lower total and pattern specific squared error loss compared to the traditionally available methods. When $Y$ is generated under the pattern mixture formulation with a MNAR missing data mechanism (MNAR PMY), PMKS and MIMI have both lower pattern specific contributions to the prediction error (PE) in the pattern where $x_1$ is missing, and lower total prediction error compared to all other methods.

As might be expected, PMKS and CCS have different out-of-sample prediction performance when the missing data are not Missing at Random (MAR). In fact, PMKS minimizes the expected prediction loss regardless of missingness mechanism, while CCS tends to rival PMKS only when the data are MAR. We will see that when the data are modified to induce a Missing Not at Random (MNAR) mechanism, PMKS has optimal predictions on average compared to traditional methods.

As both the strength of the missingness mechanism and the beta coefficient associated with the missing variable increase, the magnitude of the differences in methods favors PMKS/MIMI over all the other methods.

[Figure 2 about here.]

[Table 4 about here.]

Table 4 of out-of-sample imputations of $x_1$ provides insight into some of the biases seen in figure 2. When $Y$ is included in the imputation model during model construction, parameter estimates tend to be unbiased. However, when $Y$ is included during MI performed using predictive mean matching and chained equations, the imputations of $x_1$ show have the largest squared error of all the imputations procedures for every missing data mechanism apart from unconditional mean imputation. Although the apparent bias in imputations for missing covariates seem small, their total contribution over all individuals can be quite significant. These results show that bias in imputing missing predictors leads to poorer downstream predictions and larger prediction error for the outcome.

Papers have explored in detail the advantages of including Y in the imputation model (Moons et al., 2006). Using $Y$ in the imputation model during model construction leads to unbiased estimates of regression coefficients. Whereas this may be a fine approach during the model building (in-sample) population, it is not practical in the prediction setting where the outcome is unknown and would be imputed as part of the fixed imputation model. When using chained equations imputation model in which the covariate with the least amount of missing data (in these simulations $Y$) is imputed first, can lead to biased imputations for the next missing covariates of the chain (in these simulations $X_1$). We do not present the situation in which $Y$ was used in the in-sample imputation model to produce unbiased regression estimates, but not included in the out-of-sample imputation model - a combination which would have less propagated imputation bias. Even though it may seem that the inclusion of $Y$ in the imputation model will lower prediction error, careful thought and attention need to be placed on the practicality of this, as well as the statistical implications.

## 5. Application: SUPPORT Data Example

The Study to Understand Prognoses and Preferences for Outcomes and Risks of Treatments (SUPPORT) was a multi-center, two phase study, of 9105 patients. The primary goal of the study was to estimate survival over a 180-day period for seriously ill hospitalized adults

(Phillips et al., 1995). A component of the SUPPORT prognostic model was the SUPPORT day 3 Physiology Score (SPS), and obtaining valid predictions from the SPS Model is vital since it is the single most important prognostic factor in the SUPPORT model. The SUPPORT physiology score can range from 0 to 100 and we include the following covariates: Partial pressure of oxygen in the arterial blood, Mean blood pressure, White blood count, Albumin, APACHE III respiration score, temperature, Heart rate per minute, Bilirubin, Creatinine, and Sodium.

After excluding one individual missing SPS score, and one individual missing all covariates, 9103 individuals remained in our SPS model of which 3842 had complete data, 2323 were missing Partial pressure of oxygen in the arterial blood, 212 were missing white blood count, 3370 were missing albumin, 2599 were missing bilirubin, and 66 were missing creatinine, resulting in 23 observed missing data patterns, and 1024 possible missing data patterns. Ten-fold cross-validation was used to compare the squared error loss of MI, CCSM, MIMI and PMKS within missing data patterns, as well as total average squared error loss, weighted by proportion of individuals in each pattern.

### 5.1 *SUPPORT Example Results*

For each method, ten-fold cross validation of the prediction models was implemented. For the patterns with less than or equal to $N = (p + 1) * 2 = 22$ subjects, the complete case submodel was used, and the hybrid PMKS/CCS approach (as described in section 3.2) was implemented. For the original SUPPORT data, all methods performed similarly both across and within patterns. In our simulation, we saw similar results when data were MAR, giving rise to the possibility that these data also follow a MAR mechanism. To exaggerate the missing data mechanism, we induced a MNARY mechanism by adding 25 units to individuals SPS scores who were missing the covariate partial pressure of oxygen in the arterial blood (pafi). This resulted in a large reduction in PE under PMKS compared to traditional MI methods and CCS, for the patterns in which partial pressure of oxygen in the arterial blood was missing.

[Figure 3 about here.]

The original data results are shown in the two sub-figures in the left of figure 3. The total model PE does not differ between the four methods. When a MNAR mechanism is induced in the support data, as shown in the two left sub-figures, PMKS and MIMI outperform CCS and MI. In the patterns for which partial pressure of oxygen in the arterial blood (pafi) is missing, the benefits of PMKS and MIMI compared to CCS and MI are apparent. For these patterns, both the unweighted PE (figure 3 top-right) and weighted PE (figure 3 bottom-right) show this reduction in PE. The model PE, which is the sum of all the pattern specific contributions to the PE results in approximately a 50% reduction in PE for PMKS/MIMI compared to MI, and a 40% reduction in PE for PMKS/MIMI compared to CCS.

## 6. Remarks

Statistical literature abounds with imputation methods for model inference, but there are very few practical solutions for obtaining predictions for new individuals who do not present with all of the necessary predictors. In this paper, we have shown that PMKS provides competitive if not optimal predictions for a variety of missing data mechanisms, and has large gains in computation time since external data and imputation models are no longer needed to make new predictions. Our method is robust, straightforward to implement, and can easily be extended for any generalized linear model and models with nonlinear effects (this is ongoing work).

While PMKS is clearly optimal in the sense of minimizing an expected prediction loss, the procedure also has implications for model inference under non-MAR scenarios. It leads to an important extension of classical MI procedures for estimation where the MAR can be relaxed or assessed. Moreover, PMKS is more computationally efficient thanMI procedures. In the age of big data, this is an important consideration and driving factor in most scientific contexts with big data.

## 6.1 Remark A: Conditioning Y on X and M

One might ask whether we are interested in the model marginalized over $\boldsymbol{M}$, $E[Y|\boldsymbol{X}] = \boldsymbol{X}\boldsymbol{\beta}$, or the conditional model, $E[Y|\boldsymbol{X}, \boldsymbol{M}] = \boldsymbol{X}\boldsymbol{\beta} + \boldsymbol{M}\boldsymbol{\delta}$. This is a philosophical question with many differing viewpoints. For inferential purposes (which we do not consider here) it has been argued that the marginal model is the model of interest, however in many situations can be imagined in which the conditional model is the simpler way to express a complicated marginal model. From the prediction point of view, PMKS/MIMI model is robust to both situations and therefore is the preferred method to use.

By assuming the marginal model is correct, you are making the assumption that data are MAR and the $\boldsymbol{\delta}$ parameters are zero. As the number of covariates increase, this assumption in practice seems to be more plausible. A way to asses this in practice is to evaluate whether the MI model and PMKS/MIMI model give similar results, as we did in the SUPPORT example.

## 6.2 Remark B: The Relationship between Y and M

The relationship between $Y$ and $M$ plays an important role in our modeling assumptions. An outcome generated from a selection model formulation is assumed to be independent of the missing data mechanism, such that the outcome would be the same regardless of whether covariate information is missing or observed. The pattern mixture model formulation assumes that the missing data mechanism is part of the response model, such that the outcome can be depend on the missing data pattern. Both mechanisms are thought to exist in Biomedical data, but unfortunately the two approaches represent fundamentally different descriptions of the underlying natural process. The two approach will only coincide when $\delta_1 = ... = \delta_p = 0$ (the MAR assumption). The $\boldsymbol{\beta}$ parameters could be adjusted to account for the differential missingness across the two approaches, but then the model are not comparable.

### 6.3 *Remark C: Extending to Generalized Linear Models and Other Prediction Approaches*

We performed the same set of simulations assuming a true logistic regression model, where we used a logarithmic scoring rule to compare methods. The general ordering of results holds and will be explored in future papers. These results extend to random forests as well. For example, our results suggest a random forest should be fit by pattern, using only the data in that pattern.

Care should be taken when developing clinical prediction models when missing data is present. Prediction models should be fit with PMKS and estimation should be conducted under MIMI for optimal results, since these methods are robust to a wide variety of missing data mechanisms compared to commonly available MI and CCS methods.

### Acknowledgments

The authors wish to thank Professors Robert Greevy, Matt Shotwell, Thomas Stewart, Melinda Aldrich, Frank Harrell and Nathaniel Mercaldo for critical reading, helpful suggestions, and valuable feedback of the original version of the paper. In addition, the authors would like to thank Eric Grogan, and the TREAT Laboratory via funding from the Department of Veterans Affairs Career Development Award (10-024), and Melinda Aldrich via funding from NIH/NCI 5K07CA172294, as well as the Department of Thoracic Surgery for supporting this work. *Conflict of Interest*: None declared.

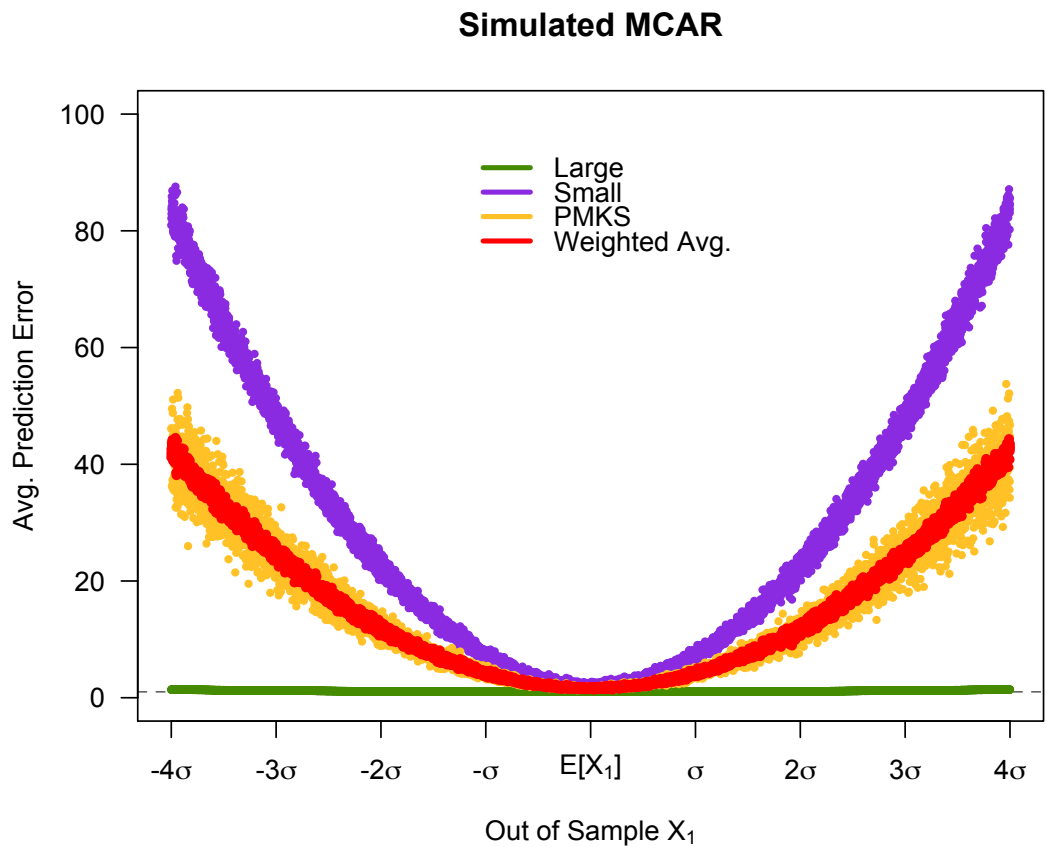

**Figure 1.** Comparison of Expected Prediction Error for the Large fully specified model: $E[Y|X_1, X_2] = \beta_0 + \beta_1 X_1 + \beta_2 X_2$, and Small underspecified model: $E[Y|X_1] = \beta_{0,3} + \beta_{1,3}^* X_1$. Pattern Mixture Kernel Submodel (PMKS) predictions are a weighted average of the Large and Small models, weighted by P(M=1)

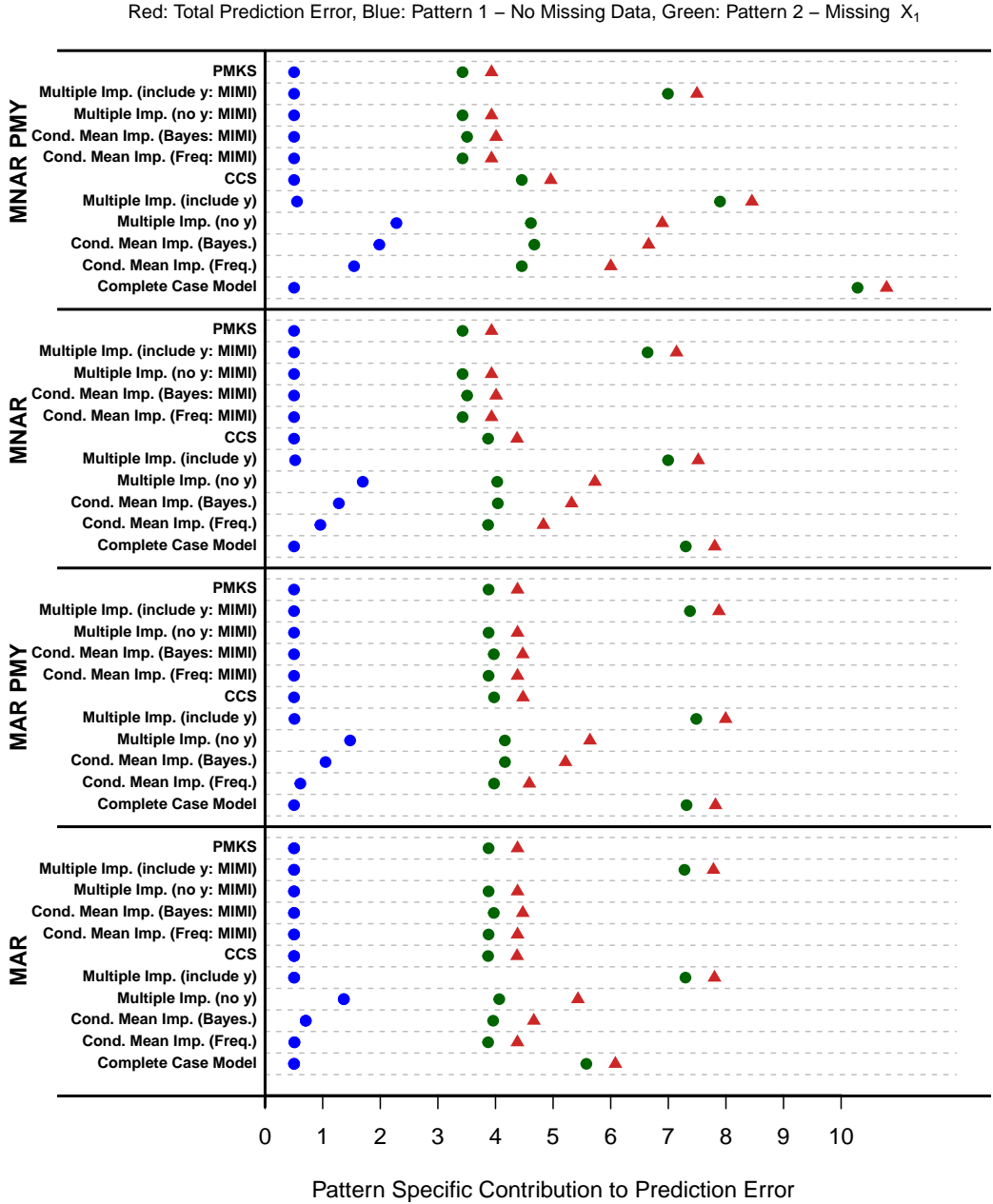

**Figure 2.** Simulation Results set of parameters: $\beta_0 = 1, \beta_1 = 3, \beta_2 = 1, \delta_1 = 1, \delta_3 = 1, P(M_1 = 1) = 0.5, \nu_1 = 1, \nu_2 = 1, \nu_{1,Y} = 1, \nu_{2,Y} = 1$. The missing data mechanisms Missing at Random (MAR) and Missing Not at Random (MNAR) were generated under a Pattern Mixture $Y$ (PMY) and Selection Model $Y$ formulation. Red triangles represent the Total Prediction Error (TPE) summed over all missing data patterns. Blue circles represent the PE for pattern 1 where there is no missing data. Green circles represent the PE for pattern 2 in which $x_1$ is missing.

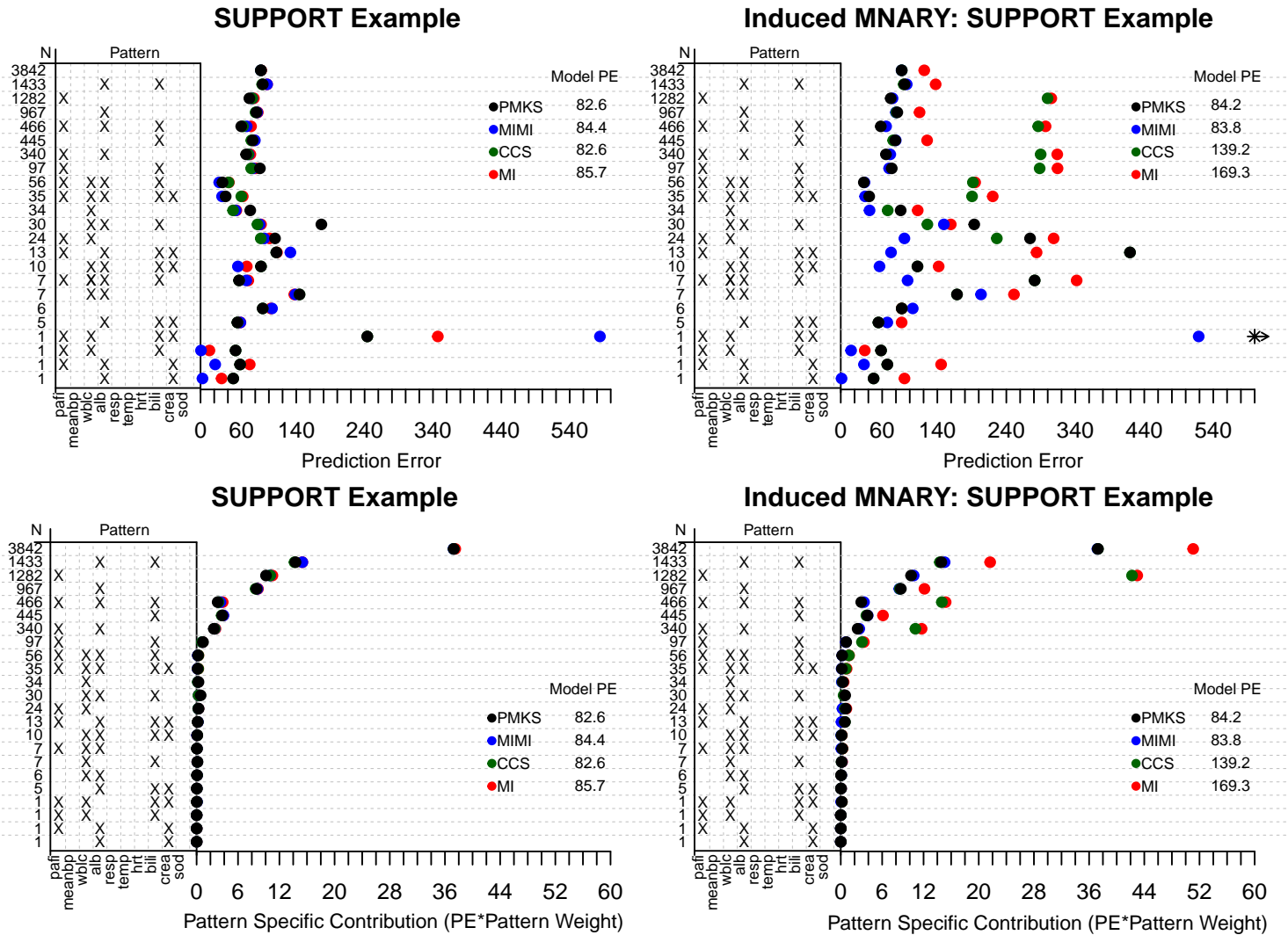

**Figure 3.** The covariates included in the SPS prediction model include Partial pressure of oxygen in the arterial blood (pafi), Mean blood pressure (meanbp), White blood count (wblc), Albumin (alb), APACHE III respiration score (resp), temperature (temp), Heart rate per minute (hrt), Bilirubin (bili), Creatinine (crea), and Sodium (sod). There are 23 patterns present in the SUPPORT data, and missing covariates are denoted with 'X'. *N* is the total number of subjects in each missing data pattern. Pattern Mixture Kernel Submodels (PMKS), Multiple Imputation with Missingness Indicators (MIMI), Complete Case Submodels (CCS), and traditional Multiple Imputation (MI) methods are all compared. The top two figures are the unweighted pattern specific PE, and the bottom two figures are the pattern specific contribution to the PE in which the partial PE is weighted by the observed proportion of individuals in each pattern.

**Table 1**

*Comparing imputation methods for an out-of-sample individual with missing data.*

| | Out-of-sample Imp. Requires | Pros | Cons |
|---|---|---|---|
| Zero Imputation | Nothing | Neglible computation time | Zero may not be an appropriate value |
| | | | Probably results in incorrect predictions |
| Mean Imputation | Unconditional means | Neglible computation time | Only works for the average individual |
| Conditional Mean Imputation | Conditional mean imputation model for every missing data pattern | Lower computation time | Large bias/variance tradeoff for MNAR |
| | | Can approximate a MI procedure | |
| CCS | Submodels to be fit | Negligible computation time | Large bias/variance tradeoff for MNAR |
| | | May be advantageous if data are MAR | |
| | | Fittable for unobserved patterns | |
| PMKS | Submodels to be fit | Negligible computation time | May be less efficient if data are MAR |
| | | Works for any missingness mechanism | Patterns with low membership may not fit well |
| MMI | Original data/Conditional distribution | Works for any missingness mechanism | High computational cost |
| | Computer/Imputation engine | Allows for efficient parameter estimation | Not viable in the clinic |
| | Original data/Conditional distribution | Established method | High computational cost |
| MI | Computer/Imputation engine | Works when data are MAR | Not viable in the clinic |
| | | | Large bias/variance tradeoff for MNAR |

**Table 2**

*Comparison of Pattern Mixture Kernel Submodels (PMKS) and Complete Case Submodels (CCS).*

| Pattern | PMKS ($\hat{f}_m$) | CCS ($\hat{g}_m$) |
|---|---|---|
| 1:$X_1^{obs}, X_2^{obs}$ | $E[Y|X_1, X_2, M_1 = 0, M_2 = 0] = \gamma_{0,1} + \gamma_{1,1}X_1 + \gamma_{2,1}X_2$ | $E[Y|X_1, X_2] = \beta_{0,1}^* + \beta_{1,1}^*X_1 + \beta_{2,1}^*X_2$ |
| 2:$X_1^{miss}, X_2^{obs}$ | $E[Y|X_2, M_1 = 1, M_2 = 0] = \gamma_{0,2} + \gamma_{2,2}X_2$ | $E[Y|X_2] = \beta_{0,2}^* + \beta_{2,2}^*X_2$ |
| 3:$X_1^{obs}, X_2^{miss}$ | $E[Y|X_1, M_1 = 0, M_2 = 1] = \gamma_{0,3} + \gamma_{1,3}X_1$ | $E[Y|X_1] = \beta_{0,3}^* + \beta_{1,3}^*X_1$ |
| 4:$X_1^{miss}, X_2^{miss}$ | $E[Y|M_1 = 1, M_2 = 1] = \gamma_{0,4}$ | $E[Y] = \beta_{0,4}^*$ |
| | $\gamma_{p,m}, \beta_{p,m}^*$ represents the effect of the $p^{th}$ covariate in pattern $m$ | |

**Table 3**

*Missing data mechanisms used for simulation. $\nu_0$ is empirically calculated to allow the probability of missingness to maintain the desired level. $expit = \frac{e^x}{1+e^x}$.*

| | **Missing Data Mechanism for $X_1$** |
|---|---|
| **MCAR** | $P(M) = expit(\nu_0)$ |
| **MAR** | $P(M) = expit(\nu_0 + \nu_2 X_2)$ |
| **MARY** | $P(M) = expit\left(\nu_0 + \nu_{2,Y}\left(\frac{Y/\sigma_y + X_2}{\sqrt{2(1+cor(Y,X_2))}}\right)\right)$ |
| **MNAR** | $P(M) = expit(\nu_0 + \nu_1 X_1)$ |
| **MNARY** | $P(M) = expit\left(\nu_0 + \nu_{1,Y}\left(\frac{Y/\sigma_y + X_1}{\sqrt{2(1+cor(Y,X_1))}}\right)\right)$ |

**Table 4**

*Squared Imputation Error of the true $X_1$ compared to the imputed $X_1$ under different imputation methods and missing data mechanisms: Imputation Error of $X_1 = \sum_i (X_{1i} - \hat{X}_{1i})^2$*

|                      | MAR  | MNAR | MAR PMY | MNAR PMY |
|----------------------|------|------|---------|----------|
| Unconditional Mean   | 0.56 | 0.56 | 0.76    | 0.76     |
| Cond. Mean           | 0.38 | 0.38 | 0.53    | 0.53     |
| Cond. Mean (Bayes)   | 0.49 | 0.49 | 0.69    | 0.69     |
| MI (no y)            | 0.47 | 0.47 | 0.61    | 0.61     |
| MI (incude y)        | 0.76 | 0.74 | 0.75    | 0.71     |